\documentclass[twocolumn,nobalancelastpage,amsmath,amssymb,nofootinbib]{revtex4}
\usepackage{graphicx}
\usepackage{bm}
\usepackage{bbm}
\usepackage[T1]{fontenc}
\usepackage[utf8]{inputenc}
\usepackage{amsmath}

\def\beq{\begin{equation}}
\def\eeq{\end{equation}}
\def\bea{\begin{eqnarray}}
\def\eea{\end{eqnarray}}

\def\r{\bm{r}}
\def\rp{\bm{r}^\prime}

\def\jp{j^\prime}

\def\s{\sigma}
\def\sp{\sigma^\prime}

\def\c{\dagger}

\begin{document}
\date{\today}

\title{
Nonequilibrium Green's function approach to multi-band Cooper-pair transport:\\
linear magnetoresistance effect due to nonunitary superconductivity
}

\author{G. Tkachov}

\affiliation{
Institute of Physics, Augsburg University, 86135 Augsburg, Germany}

\begin{abstract}
Many-body transport has emerged as an efficient tool for understanding interaction effects in quantum materials with a multi-band electronic structure. 
This paper proposes a formula for the two-particle transmission coefficient for Cooper-pair transport between multi-band normal and superconducting materials.
The approach employs a tight-binding nonequilibrium Green's function technique, allowing a direct calculation of the two-particle current, without invoking the paradigm of Andreev reflection. As an application of the theory, we demonstrate a low-field linear magnetoresistance effect for superconductors with an induced nonunitary order parameter. These results uncover an unexplored route for detecting unconventional nonunitary superconductivity in quantum materials of current theoretical and experimental interest.
\end{abstract}

\maketitle

\section{Introduction}

Quantum materials are a burgeoning frontier of condensed matter physics and materials science \cite{Keimer17,Giustino20}.
In such materials, quantum effects emerge at macroscopic scales as a result of reduced dimensionality, interactions, band structure specifics or 
a combination of these factors. The recent flurry of results on topological insulators \cite{Liu16}, Weyl semimetals \cite{Armitage18}, 
transition metal dichalcogenides (TMDs) \cite{Choi17}, topological and unconventional superconductors \cite{Tanaka12,Sato17,Ghosh20,Culcer20,Nakai21,Cheng21,Ohashi21,Ramires22} is a case in point. 
A common thread that links these groups of materials is their multi-band electronic structure associated with internal degrees of freedom (DOF). 
These include the electron spin and, depending on the context, atomic orbitals, crystal sublattices or/and the particle-hole degree of freedom.  

The purpose of this paper is twofold. First, we revisit some many-body aspects of electron transport theory for multi-band materials. 
On the one-particle level or within one-particle approaches, the progress in understanding multi-band transport properties is significant 
(see e.g. review \cite{Culcer20} and references therein). 
Many-body transport, on the other hand, has been explored to a much lesser extent even at the two-particle level. 
A typical and important example is the transport of Cooper pairs across an interface between normal and superconducting materials.
The Cooper-pair transport offers insights into the superconducting order parameter and is traditionally 
treated in the language of Andreev reflection \cite{Blonder82}.
For multi-band materials, however, the number of channels available for Andreev reflection can be very large.
First of all, because of the spin degree of freedom, two electrons in a Cooper pair can form one singlet and three triplet states, 
which yields a total of four possible scattering channels. 
If, additionally, the pairing involves multiple orbitals or sublattices, the number of the scattering channels increases greatly, 
e.g. for two atomic orbitals (acting as pseudospin) it jumps to sixteen since the pseudospin singlet and triplets come now into play as well. 
A large number of transport channels makes the problem intriguing, 
but keeping track of them in the Andreev reflection language becomes increasingly difficult.

Instead, in this paper we directly compute the two-particle transmission coefficient for Cooper pairs, using the tight-binding approach in combination with the nonequilibrium Green's function (NEGF) technique. The calculation builds upon the original proposal of Ref. \cite{Cuevas96} and more recent work \cite{GT17}, taking into account the lattice structure, multiple internal DOF and pairing states as well as the position dependence of real-space NEGFs. 
We obtain the pair transmission in a compact analytic form applicable to diverse multi-band materials interfaces. 

Another goal of this paper is to examine the pair transport for unconventional triplet superconductors with broken time-reversal symmetry (TRS).
Such superconductors can often be identified as nonunitary superconductors and are currently the subjects of intense experimental and theoretical effort in a broader context of quantum materials (see e.g. \cite{Hillier09,Hillier12,Lado19,Shang22,Ghosh22,Wolf22} and reviews \cite{Ghosh20,Culcer20,Ramires22}). 
From the perspective of a transport study, it is convenient to define unitary and nonunitary superconducting states, 
using the matrix condensate Green’s function $\hat{F}$.\footnote{This definition (see e.g. \cite{Culcer20}) is complementary to that using the gap function, cf. \cite{Sigrist91} and recent reviews \cite{Ghosh20,Ramires22}.} 
The state is then called unitary if $\hat{F}\hat{F}^\c$ is proportional to a unit matrix, otherwise it is called nonunitary. For example, in the singlet-triplet basis
$
 \hat{F} = \hat{f}_0 i\sigma_y + \hat{\bm f}\cdot {\bm \sigma}i\sigma_y,
$
where ${\bm \sigma}$ is the vector of Pauli matrices $\sigma_x,\sigma_y$ and $\sigma_z$, 
while $\hat{f}_0$  and $\hat{\bm f}=[\hat{f}_x,\hat{f}_y,\hat{f}_z]$ are the amplitudes of the singlet and triplet pairing, being matrices in all DOF other than 
spin.\footnote{Hereafter, direct products of the $f$- and $\sigma$-matrices are assumed. The “dot” and “cross” mean the usual scalar and vector products.} 
We have then

\beq
\hat{F}\hat{F}^\c = (\hat{f}_0\hat{f}^\c_0 + \hat{\bm f} \cdot \hat{\bm f}^\c ) \sigma_0 + 
(\hat{f}_0 \hat{\bm f}^\c + \hat{\bm f} \hat{f}^\c_0 + i\hat{\bm f} \times \hat{\bm f}^\c) \cdot {\bm \sigma},
\label{FF}
\eeq
where the first and second terms correspond to unitary and nonunitary pairing, respectively, in spin space ($\sigma_0$ is the unit spin matrix).
In particular, the axial vector $i\hat{\bm f} \times \hat{\bm f}^\c$ indicates the TRS breaking due to emergent magnetization of the triplet pairs. 

Beyond the single-orbital picture, one encounters a bigger family of the candidate nonunitary order parameters (see e.g. \cite{Ghosh20,Ramires22,Wolf22}),
so the ability to access and characterize them becomes instrumental.
Experimentally, the states with broken TRS can be detected by using the muon spin relaxation technique \cite{Ghosh20}.
Theoretical efforts to identify nonunitary states in observables have been focused on the electric conductance of normal-superconductor interfaces 
\cite{GT17,Honerkamp98,Linder07,GT19} and on finite-energy gap structures in the spectrum of superconductors \cite{Ramires22,Lado19}. 

\begin{figure}[t]
\begin{center}
\includegraphics[width=90mm]{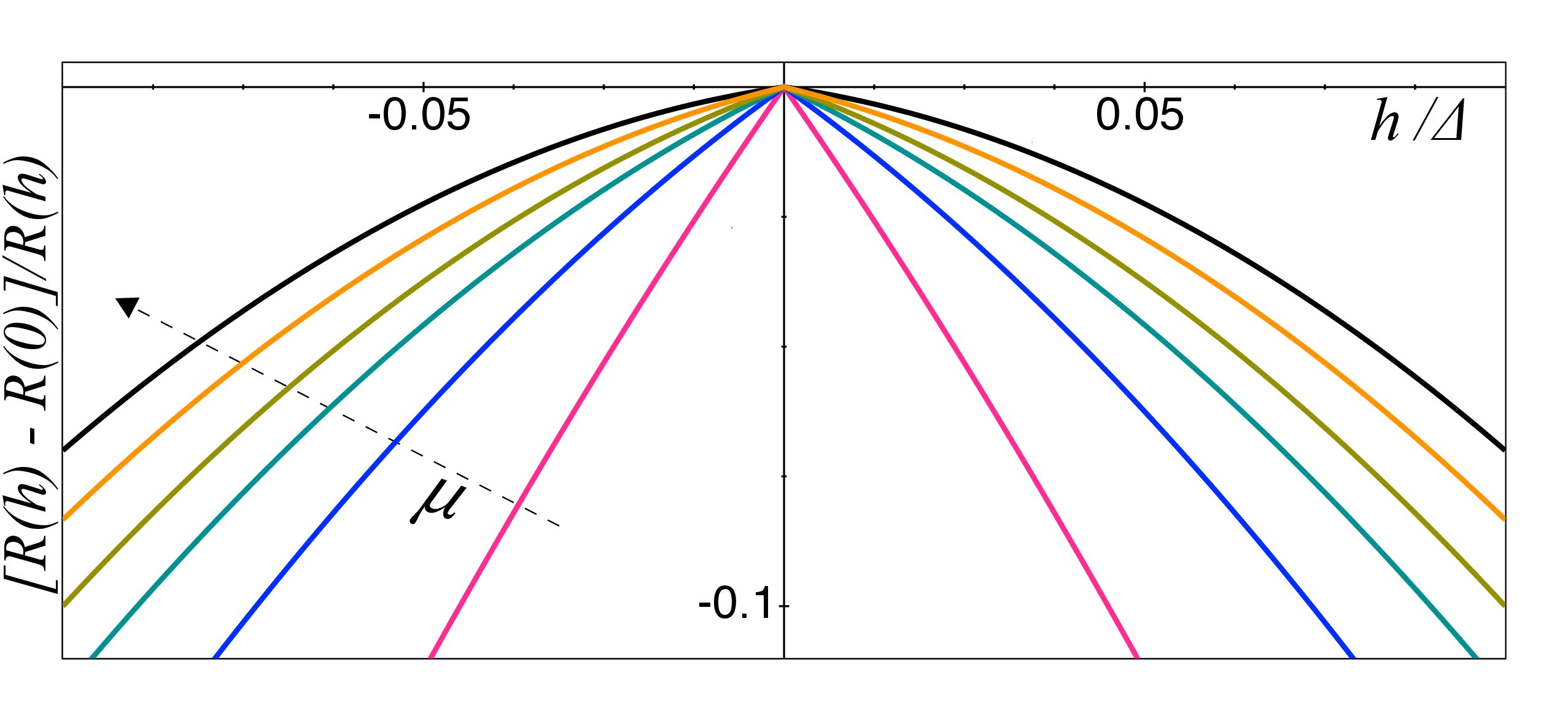}
\end{center}
\caption{
Interface magnetoresistance (\ref{dR}) versus Zeeman splitting in the superconductor.
The linear magnetoresistance for $|h| \ll \Delta$ indicates a TRS-breaking nonunitary triplet state.
The chemical potential increases from $\mu = \Delta$ to $\mu = 10 \Delta$ as shown by the arrow (see also Eqs. \ref{alpha} and \ref{beta}).
}
\label{dR_fig}
\end{figure}

Generally, signatures of nonunitary states remain elusive if there is no means of controlling the TRS breaking in the superconducting state.
In that regard, it would be helpful to identify a magnetotransport effect associated with and controllable through the axial vector $i\hat{\bm f} \times \hat{\bm f}^\c$.
We propose a prototype of such an effect in a bilayer system comprising a spin-orbit-coupled superconductor and 
a quantum anomalous Hall insulator (QAHI). 
In this case, the nonunitarity emerges as the combined effect of spin-orbit coupling (SOC) and Zeeman field, 
${\bm h}$, caused by the magnetization of the QAHI layer. Furthermore, the QAHI acts as a spin filter, enabling the observation of
the triplet pair magnetoresistance which we find to be

\beq
\frac{R(h) - R(0)}{R(h)} = -\alpha \frac{|h|}{\Delta} \left( \beta + \frac{|h|}{\Delta} \right).
\label{dR}
\eeq
Above, $h$ is the projection of ${\bm h}$ on the interface normal, $\Delta$ is the energy gap of the superconductor, 
and $\alpha$ and $\beta$ are material parameters. 

A distinct feature of Eq. (\ref{dR}) is a linear low-field dependence on $|h|$ (see also Fig. \ref{dR_fig}). 
It indicates the broken TRS associated with the axial vector $i\hat{\bm f} \times \hat{\bm f}^\c \propto {\bm h}$ and, therefore, 
provides direct evidence for nonunitary triplet pairing.
Further details of the model and calculation as well as a comprehensive discussion of the results are given in the sections below.

\section{NEGF technique for multi-band interface transport}

We begin by briefly outlining the NEGF method for interface transport. 
The goal is to introduce matrix notations that enable efficient computation of the interface current for multi-band materials.

\subsection{Interface Hamiltonian and current operator}
\label{Model}

\begin{figure}[t]
\begin{center}
\includegraphics[width=80mm]{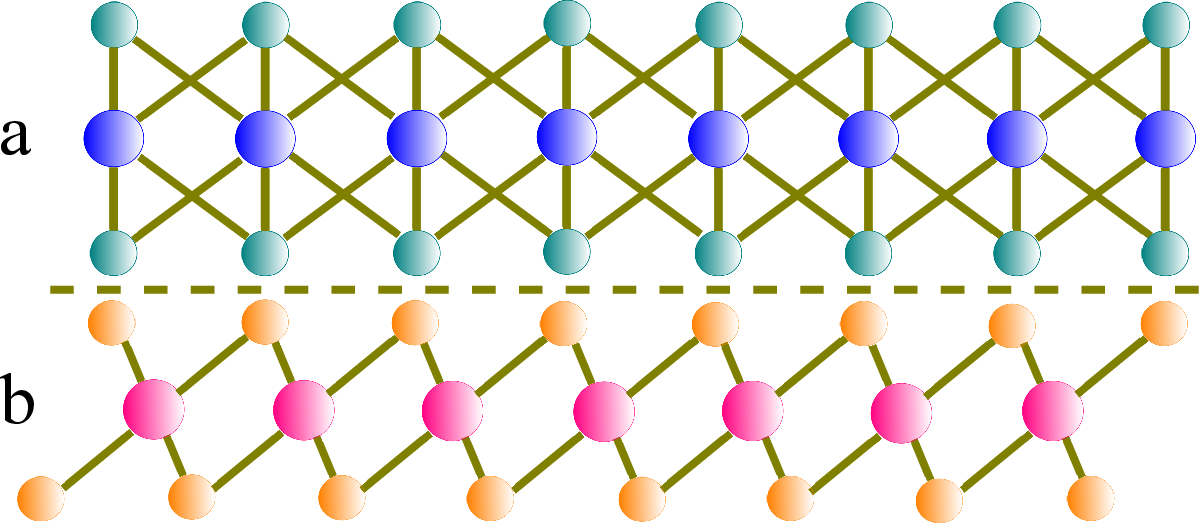}
\end{center}
\caption{
Interface between two layered TMDs as an example of a multi-band materials system 
for NEGF transport calculation (see Sec. \ref{Model}). 
The tight-binding model accounts for internal DOF consisting of the spin states and atomic orbitals 
(for $d$ - electrons in the case of TMDs).
}
\label{Interface}
\end{figure}

We consider two generic crystalline materials (here $a$ and $b$) 
forming a planar interface (see also Fig. \ref{Interface}). The structure is described by the Hamiltonian 
$
H = H_a + H_b + H_{ab},
$
where $H_a$ and $H_b$ are the tight-binding Hamiltonians of the constituents, while $H_{ab}$ accounts for tight-binding hopping across 
the interface:

\beq
H_{ab} = \sum\limits_{\r\rp \\ \s\sp \\ j\jp} 
\left[ b^\c_{\r \s j} \left(t_{\r\rp}\right)^{\s\sp}_{j\jp} a_{\rp\sp\jp} + h.c. \right].
\label{H_ab}
\eeq
Operator $a_{\r \s j}$ ($a^\c_{\r \s j}$) annihilates (creates) an electron in material $a$ on lattice site $\r$ with spin projection $\s=\uparrow,\downarrow$ 
in atomic orbital $j=1,2,...$; $b_{\r \s j}$ and $b^\c_{\r \s j}$ are the analogous operators for material $b$, and 
$\left(t_{\r\rp}\right)^{\s\sp}_{j\jp}$ is the hopping energy between the states with $\r \s j$ and $\rp\sp\jp$.
Correspondingly, the operator of the electric current between $a$ and $b$ is given by
 
\beq
I = \frac{ie}{\hbar} \sum\limits_{\r\rp\s\sp j\jp} \left[ b^\c_{\r \s j} \left(t_{\r\rp}\right)^{\s\sp}_{j\jp} a_{\rp\sp\jp} - h.c. \right].
\label{I_ab}
\eeq

We proceed by recasting the interface Hamiltonian into a compact quadratic form,

\beq
H_{ab} = \frac{1}{2} \sum\limits_{\r\rp} C^\c_{\r} \hat{V}_{\r\rp} C_{\rp},
\label{H_ab_C}
\eeq
using matrix notations for electron operators $C_{\r}$, $C^\c_{\r}$, and hopping energy  $\hat{V}_{\r\rp}$  
which absorb all discrete DOF, i.e. the material constituents ($a, b$), particle and hole states (p,h), spin species ($\uparrow,\downarrow$) and atomic orbitals ($1,2,...$). The hierarchy of the DOF is chosen as follows. First, $C_{\r}$ is two-component in $ab$ space:

\beq
C_{\r} = 
\left[
\begin{matrix} 
A_{\r} \\
B_{\r}
\end{matrix}
\right],
\label{C}
\eeq
where operators $A_{\r}$ and $B_{\r}$ act in the subsystems. Each of these operators is, in turn, two-component in ph space:  

\beq
A_{\r} = 
\left[
\begin{matrix} 
a_{\r} \\
a^\c_{\r}
\end{matrix}
\right],
\quad
B_{\r} = 
\left[
\begin{matrix} 
b_{\r} \\
b^\c_{\r}
\end{matrix}
\right].
\label{AB}
\eeq
This (Nambu) representation applies equally to normal and superconducting materials and brings along the particle-hole symmetry 
[which explains the prefactor of $1/2$ in Eq. (\ref{H_ab_C})]. 
Further, both operators $a_{\r}$ and $b_{\r}$ are naturally two-component spinors:

\beq
a_{\r} = 
\left[
\begin{matrix} 
a_{\uparrow\r} \\
a_{\downarrow\r}
\end{matrix}
\right],
\quad
b_{\r} = 
\left[
\begin{matrix} 
b_{\uparrow\r} \\
b_{\downarrow\r}
\end{matrix}
\right],
\label{ab}
\eeq
and, lastly, $a_{\r\s}$ and $b_{\r\s}$ have the orbital components:

\bea
a_{\r\s} = 
\left[
\begin{matrix} 
a_{\r\s 1} \\
a_{\r\s 2} \\
...
\end{matrix}
\right],
\quad
b_{\r\s} = 
\left[
\begin{matrix} 
b_{\r\s 1} \\
b_{\r\s 2} \\
...
\end{matrix}
\right].
\label{ab_orb}
\eea
The hopping matrix $\hat{V}_{\r\rp}$ has the same hierarchy of the discrete DOF and 
unfolds as follows

\bea
&
\hat{V}_{\r\rp} = \left[
\begin{matrix} 
0 & \hat{T}^\c_{\r\rp} \\ 
\hat{T}_{\r\rp} & 0
\end{matrix}
\right], 
&
\label{V_ab}\\
&
\hat{T}_{\r\rp} = \left[
\begin{matrix} 
\hat{t}_{\r\rp} & 0 \\ 
0 & -\hat{t}^*_{\r\rp}
\end{matrix}
\right],
&
\label{T_ph}\\
&
\hat{t}_{\r\rp}= \left[
\begin{matrix} 
(\hat{t}_{\r\rp})^{\uparrow\uparrow} & (\hat{t}_{\r\rp})^{\uparrow\downarrow} \\ 
(\hat{t}_{\r\rp})^{\downarrow\uparrow} & (\hat{t}_{\r\rp})^{\downarrow\downarrow}
\end{matrix}
\right],
&
\label{t_spin}\\
&
\quad
(\hat{t}_{\r\rp})^{\s\sp}= \left[
\begin{matrix} 
\left( \hat{t}_{\r\rp} \right)^{\s\sp}_{11} &  \left( \hat{t}_{\r\rp} \right)^{\s\sp}_{12} & ...\\ 
\left( \hat{t}_{\r\rp} \right)^{\s\sp}_{21} & \left( \hat{t}_{\r\rp} \right)^{\s\sp}_{22} & ... \\
... & ... & ...
\end{matrix}
\right].
&
\label{t_orb}
\eea
Here, matrices (\ref{V_ab}) -- (\ref{t_orb}) act in $ab$, ph, spin and orbital spaces, respectively.
Obviously, the Hamiltonian of the constituents, $H_a+H_b$, can be written as in terms of the same operators $C_{\r}$ and $C^\c_{\r}$,
so the Hamiltonian of the entire system is 

\beq
H = \frac{1}{2} \sum\limits_{\r\rp} C^\c_{\r}\left(\hat{H}^{0}_{\r\rp} +  \hat{V}_{\r\rp} \right) C_{\rp}. 
\label{H_C}
\eeq
As the hopping between the constituents is included in $\hat{V}_{\r\rp}$,
matrix $\hat{H}^{0}_{\r\rp}$ is diagonal in $ab$ space,

\beq
\hat{H}^{0}_{\r\rp} = \left[
\begin{matrix} 
\hat{H}^a_{\r\rp} & 0 \\ 
0 & \hat{H}^b_{\r\rp}
\end{matrix}
\right],
\label{H^0}
\eeq
where $\hat{H}^{a,b}_{\r\rp}$ are matrices in the internal DOF of materials $a$ and $b$. 
Finally, in the same matrix notations the current operator in Eq. (\ref{I_ab}) takes the form

\beq
I = -\frac{ie}{2\hbar} \sum\limits_{\r\rp} C^\c_{\r} \nu_3\tau_3i_0 \hat{V}_{\r\rp} C_{\rp},
\label{I_ab_C}
\eeq
where $\nu_3$ and $\tau_3$ are Pauli matrices in $ab$ and ph spaces; $i_0$ is the unit matrix in the remaning discrete DOF, and 
the product of these three matrices is direct. 

\subsection{Nonequilibrium Green’s functions}
\label{NEGFs}

The observable current, ${\cal I}$, is defined as the expectation value of operator $I$ (\ref{I_ab_C}) over the nonequilibrium state of the system 
and can be expressed as an integral over the energy ($E$) axis:

\beq
{\cal I} = -\frac{e}{4\pi\hbar} \int 
\sum\limits_{\r\rp} 
{\rm Tr}[
\nu_3\tau_3i_0\hat{V}_{\rp\r} \hat{G}^>_{\r\rp}(E)
]dE.
\label{I_G_E}
\eeq
Here, $\hat{G}^>_{\r\rp}(E)$ is the greater function, being a matrix element of the time-ordered NEGF (see e.g. \cite{Cuevas96}):

\bea
\breve{G}_{\r\rp}(E)= 
\left[
\begin{matrix}
\hat{G}_{\r\rp}(E) & \hat{G}^<_{\r\rp}(E) \\
\hat{G}^>_{\r\rp}(E)  & \hat{\overline G}_{\r\rp}(E)
\end{matrix}
\right].
\label{G_NE}
\eea
Its elements can be expressed in terms of the advanced (A), retarded (R), and Keldysh (K) Green’s functions:

\bea
&
\hat{G} = \frac{1}{2}\left[ \hat{G}^K + \hat{G}^A + \hat{G}^R \right],
\hat{\overline G} = \frac{1}{2}\left[ \hat{G}^K - \hat{G}^A - \hat{G}^R \right],
&
\nonumber\\
&
\hat{G}^< = \frac{1}{2}\left[  \hat{G}^K + \hat{G}^A - \hat{G}^R \right],
\hat{G}^> = \frac{1}{2}\left[ \hat{G}^K - \hat{G}^A + \hat{G}^R \right].
&
\nonumber
\eea
Their arguments have been omitted for brevity.
The summations over ${\bm r}$ and ${\bm r}^\prime$ can be absorbed into the trace operation, ${\rm Tr}$, 
so the expression for the current is in fact independent of the initially chosen representation:

\beq
{\cal I} = - \frac{e}{4\pi\hbar} \int 
{\rm Tr}[
 \nu_3\tau_3i_0 \hat{V} \hat{G}^>(E)
]dE.
\label{I_G}
\eeq
The next stage is to determine the greater function $\hat{G}^>(E)$ and trace out the $ab$ and ph DOF.
These calculations (see Appendix \ref{A}) yield the interface current for the magnetotransport problem considered next.

\section{Pair magnetotransport across a QAHI/superconductor interface}

Summarizing the results of Appendix \ref{A}, we write the interface current as the sum of one-particle (${\cal I}_1$), two-particle (${\cal I}_2$) and higher-order many-particle terms: ${\cal I} = {\cal I}_1 + {\cal I}_2 +...$.  
Of particular interest is the low-energy transport inside the excitation gap, a regime in which $ {\cal I}_1$ is negligible compared to the many-particle contributions. 
We also assume a low-transparency interface\footnote{This case is relevant for order-parameter spectroscopy (e.g. \cite{Kashiwaya00}).} 
and focus on the main contribution -- the two-particle current:  

\bea
{\cal I}_2 = \frac{e}{2\pi\hbar}
\int {\cal T}_2(E)[f_{_N}(E) + f_{_N}(-E) - 1]dE.
\label{I_2_final}
\eea
Here, 
\bea
{\cal T}_2(E) = (2\pi)^2{\rm Tr}\Bigl\{
\hat{t}^\c \hat{A}(E)\hat{t} \hat{F}(E) \bigl[ \hat{t}^\c \hat{A}(-E)\hat{t} \bigr]^* \hat{F}^\c(E)
\Bigr\}\,\,\,
\label{Tr_2}
\eea
is the two-particle transmission coefficient.
It is given in terms of the condensate Green’s function of the superconductor, $\hat{F}(E)$, 
the spectral function of the normal system, $\hat{A}(E)$, and the interface hopping matrix, $\hat{t}$.
As expected, a Cooper pair is transmitted via TRS-partner states with energies $\pm E$ and occupation numbers $f_{_N}(\pm E)$ in the normal system.\footnote{We assume a small bias voltage producing 
$f_{_N}(E) \not= 1- f_{_N}(-E)$.}

We note that Eq. (\ref{Tr_2}) is valid for multi-band normal and superconducting materials with arbitrary number of discrete DOF.
The superconducting order parameter is also rather general. 
These two aspects make our approach different from the Andreev reflection paradigm.
The latter requires the construction of the scattering states, which is a rather challenging task for multi-band materials with 
a generic order parameter. For this reason, this paper employs a tight-binding nonequilibrium Green's function technique
without invoking the paradigm of Andreev reflection.

In the following, we evaluate ${\cal T}_2(E)$ for a planar interface between two-dimensional (2D) systems: 
a QAHI and a superconductor (S) with Rashba and Zeeman interactions (see also Fig. \ref{MTMD_fig}).

\subsection{QAHI: The case of a layered magnetic TMD}
\label{QAHI}

The advantage of a QAHI is that its spectral function $\hat{A}(E)$ is robust against disorder because it is 
determined by a chiral edge state \cite{Liu16}.  As a suitable candidate material we choose a layered TMD 
because it is effectively 2D and its band-structure near the Fermi level can be modelled by a four-band Hamiltonian.
We start with the Hamiltonian for a nonmagnetic TMD (see e.g. \cite{Shi19}) and add the 
Zeeman interaction to it as follows

\bea
&&
\hat{H} = 
(\epsilon_0 -  \epsilon_x \partial^2_x +\epsilon_y k^2) s_0 \sigma_0  
- iv_x \partial_x s_x \sigma_z - v_y k s_y \sigma_0  + 
\nonumber\\
&&
(\delta_0 -  \delta_x \partial^2_x + \delta_y k^2) s_z \sigma_0 - J s_0\sigma_z - m s_z\sigma_z.
\label{H_MTMD}
\eea
This Hamiltonian acts in the space of two orbital and two spin states, where Pauli matrices $s_i$ and $\sigma_i$ (with $i=x,y,$ or $z$) 
represent the orbital and spin DOF, respectively; $s_0$ and $\sigma_0$ are the corresponding unit matrices, 
$\epsilon_0$, $\delta_0$, $\epsilon_{x,y}$, $\delta_{x,y}$ and $v_{x,y}$ are the band structure parameters of the model,
while energies $J$ and $m$ characterize the Zeeman splitting, such that $2(J + m)$ and $2(J - m)$ 
are the splittings of the conduction and valence bands, respectively. 
Also, Eq. (\ref{H_MTMD}) implies that the edge is in the $y$ - direction, and $k$ is the conserved wave number.

\begin{figure}[t]
\begin{center}
\includegraphics[width=60mm]{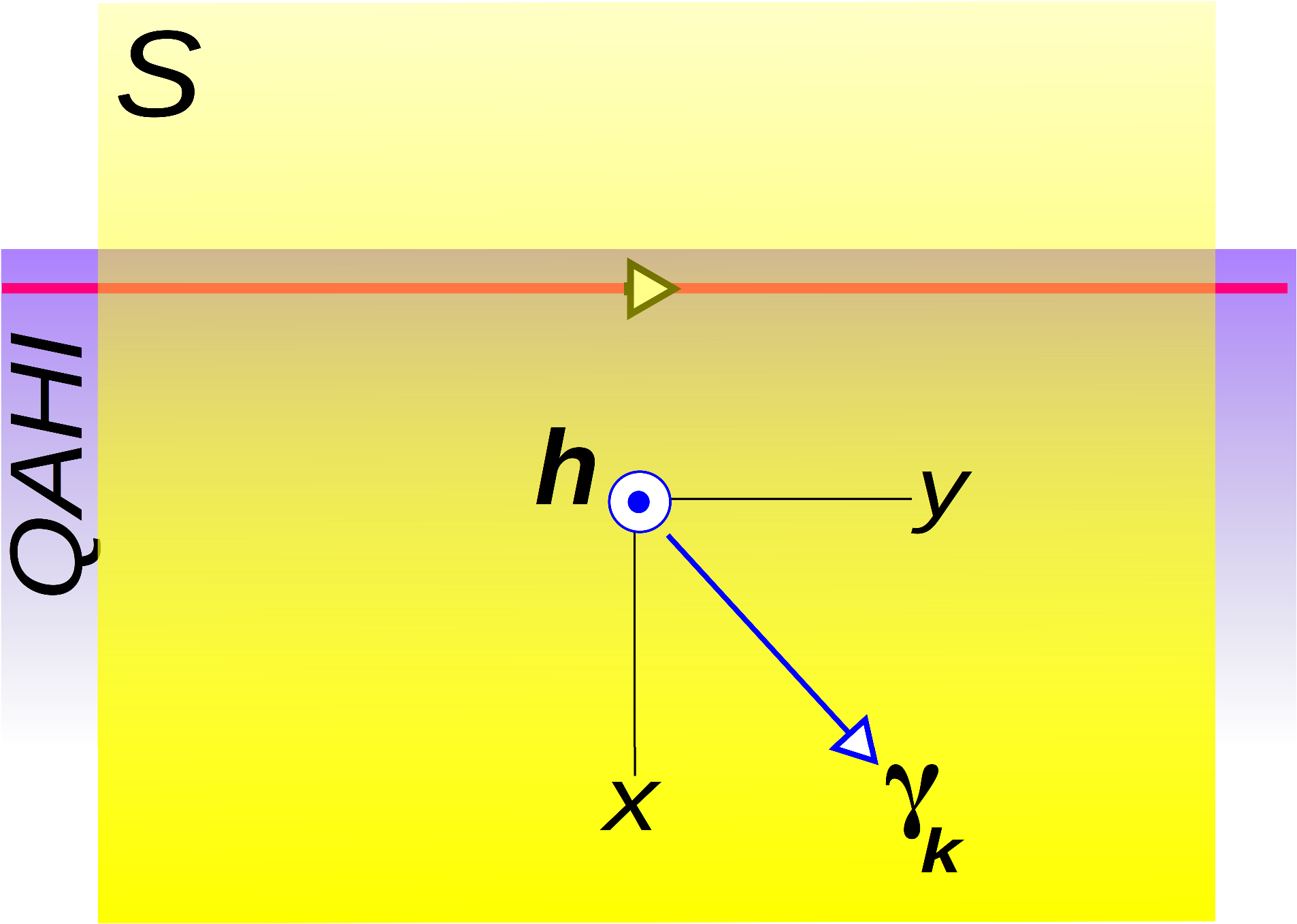}

\includegraphics[width=60mm]{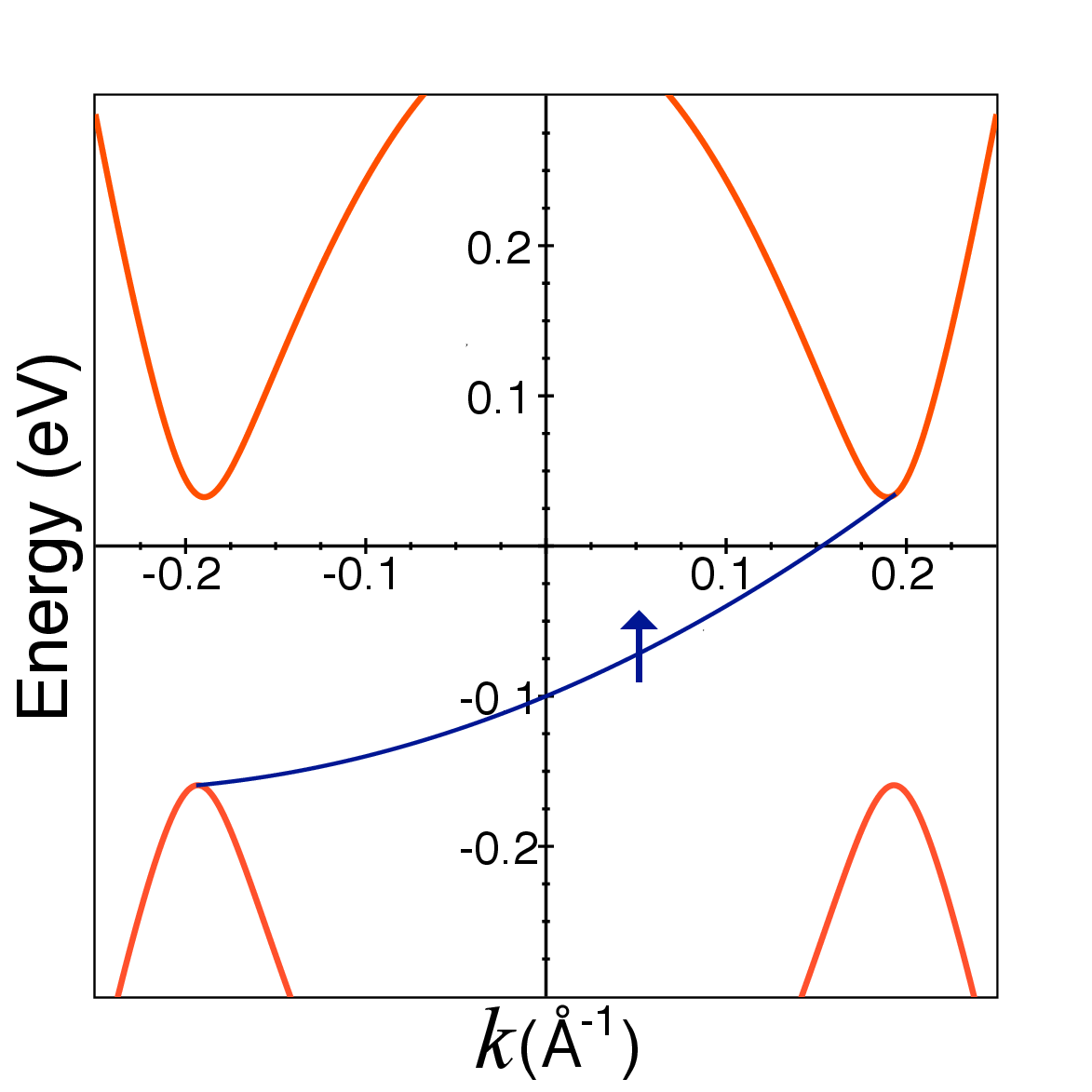}
\end{center}
\caption{
Top: Schematic of a planar interface between a QAHI (a layered magnetic TMD) and a superconducting layer (S).
Bottom: Band structure of spin-$\uparrow$ electronic states in a magnetic TMD from Eqs. (\ref{H_MTMD}) and (\ref{E_edge}).
The two bulk valleys are connected by a chiral spin-polarized edge mode crossing the Fermi level at energy $E=0$. 
The spin-$\downarrow$ states (not shown) exhibit a band gap around $E=0$ and do not contribute to transport at the Fermi level.
The plot shows the band dispersion along the $y$ direction for representative parameters
$\epsilon_0 = 0.2$ eV, $\epsilon_y= 1$ eV$\mathring{\rm A}^2$, $v_y = 0.5$ eV$\mathring{\rm A}$, $\delta_0 = -0.2$ eV, $\delta_y = 12$ eV$\mathring{\rm A}^2$, and $J = m = 0.25$ eV.
}
\label{MTMD_fig}
\end{figure}

Solving the edge problem (see e.g. \cite{Liu16,GT22}), we readily find an edge mode as
\beq
E_k = \epsilon_0 - |J|  + \epsilon_y k^2 + {\rm sgn}(m) v_y k.  
\label{E_edge}
\eeq
For $J, m > 0$ its spin is in the $z$ - direction ($\uparrow$), so is the magnetization.
In the band structure, the edge mode crosses the gap between the conduction and valence bands of the TMD, connecting its two valleys, 
as depicted in Fig. \ref{MTMD_fig}.
At the same time, all spin-$\downarrow$ states remain gapped near the Fermi level ($E=0$), so the in-gap spectral function
$\hat{A}(E)$ is determined solely by the edge mode. 
For $J, m < 0$ the magnetization is antiparallel to the $z$ - axis. The edge mode has then spin $\downarrow$ and opposite chirality ${\rm sgn}(m)=-1$. 
In both cases, $|m| > |\delta_0|$.

The knowledge of the spectrum and wave function (see \cite{Liu16,GT22}) makes the calculation of $\hat{A}(E)$ straightforward. 
The result can be written as a direct product of the matrices in coordinate, spin and orbital spaces: 

\bea
\hat{A}(E) = \hat{\rho}(E) \frac{\sigma_0 + {\bm\sigma} \cdot {\bm m}}{2} \frac{s_0 + s_y}{2}, \quad {\bm m} = {\rm sgn}(m) {\bm z},
\label{A_N_res}
\eea
where the coordinate matrix, $\hat{\rho}(E)$, is defined by its matrix elements

\bea
&&
\rho_{\bm{r}\bm{r}^\prime}(E) = \sum_k c_k(\bm{r}) c^*_k(\bm{r}^\prime)\delta(E - E_k),
\label{rho}\\
&&
c_k(\bm{r}) = \sqrt{N_k} [e^{-\varkappa_1(k)(x-x_0)} - e^{-\varkappa_2(k)(x-x_0)}] e^{iky}, \qquad
\label{c_k}
\eea
with 
$
\varkappa_{1,2}(k) = \left| \frac{ v_x }{2\delta_x }\right| \pm \sqrt{ \left( \frac{ v_x }{2\delta_x } \right)^2 + \frac{ \delta_0 - |m| + \delta_y k^2}{ \delta_x } },
$
and $N_k$ being the normalizing factor. The matrix structure of Eq. (\ref{A_N_res}) reflects the key properties of the edge mode: 
(i) the coordinate matrix $\hat{\rho}(E)$ accounts for its exponential decay from the edge (located at $x=x_0$), 
(ii) the spin matrix $\frac{1}{2}(\sigma_0 +  {\bm\sigma} \cdot {\bm m})$ is the projector on the magnetization direction 
specified by unit vector  ${\bm m}$, and 
(iii) the orbital matrix $\frac{1}{2}(s_0 + s_y)$ indicates the interband character of the edge mode 
as a mixture of the atomic orbitals of the conduction and valence bands.

\subsection{Superconductor: Nonunitarity from Rashba and Zeeman effects}
\label{S}

In the chosen geometry, the QAHI contacts mainly the interior of the S layer, which 
minimizes the role of boundary effects in S (see Fig. \ref{MTMD_fig}). 
We can therefore use a Bogoliubov-de Gennes (BdG) Hamiltonian in a 2D ${\bm k}$ space: 

\beq
H_{BdG}= 
\left[
\begin{matrix}
{\cal H}_{\bm k} & \Delta i\sigma_y \\
-\Delta i\sigma_y & -{\cal H}^*_{-{\bm k}}
\end{matrix}
\right],
\qquad 
\bm{k} = [k_x, k_y, 0],
\label{H_BdG}
\eeq
where $\Delta$ is an $s$-wave pair potential (assumed real valued), while

\beq
{\cal H}_{\bm k}=(\varepsilon_0 + \varepsilon_1 {\bm k}^2) \sigma_0s_0 + (\lambda_0 + \lambda_1{\bm k}^2)\sigma_0s_z + (\bm{\gamma}_{\bm k} - \bm{h})\cdot \bm{\sigma}s_0 
\label{H_S}
\eeq
is the four-band normal-state Hamiltonian of the layer, in which orbital matrices $s_z$ and $s_0$ act in the conduction-valence band space, 
and $\bm{\gamma}_{\bm k}$ and $\bm{h}$ are the Rashba and Zeeman fields, respectively, given by 
\beq
\bm{\gamma}_{\bm k} = \alpha_{\rm so} [k_y, -k_x, 0], \qquad \bm{h} = [0,0,h].
\label{SOC_Z}
\eeq
Above, $\alpha_{\rm so}$ is the SOC constant, while parameters $\varepsilon_0, \varepsilon_1, \lambda_0$ and $\lambda_1$ 
(all positive with $\lambda_1 > \varepsilon_1$) yield the positions and curvatures of the conduction and valence bands.  
Typically, SOC and Zeeman terms are both smaller than $\Delta$, 
and in the first approximation the pair potential in Eq. (\ref{H_BdG}) remains unaffected by the spin interactions.
Also, since the QAHI layer is magnetized out of plane, the choice of the perpendicular $\bm{h}$ is justified, 
albeit in-plane magnetic fields offer an interesting alternative that has been explored recently in the context of topological superconductivity \cite{Loder13,Loder15}.

Here, we need only the condensate Green’s function, $\hat{F}(E)$.
It can be obtained by solving the BdG equation for the Green’s function in ph space, a standard calculation (see e.g. \cite{GT17,GT19}) 
which yields

\begin{equation} 
\hat{F}(E) = [\hat{f}_0(E)i\sigma_y + \hat{{\bm f}}(E) \cdot {\bm \sigma} i\sigma_y] \frac{s_0 + s_z}{2}.
\label{F_S}
\end{equation}
We leave the technical details aside, focusing rather on the physical picture behind this result.

It is assumed that the Fermi level lies in the conduction band and the pairing affects only the conduction-band states. 
The latter assumption is justified since the pairing energy $\Delta$ is typically much smaller than the band gap $2\lambda_0$. 
This explains the presence of projector $\frac{1}{2}(s_0 + s_z)$ on the conduction band in the orbital space.
Further, the brackets in Eq. (\ref{F_S}) enclose the singlet and triplet pair functions, which are the direct products of the spin matrices and
matrices $\hat{f}_0(E)$ and $\hat{{\bm f}}(E)$ in ${\bm k}$ space, e.g. 

\bea
&&
{\bm f}_{{\bm k} {\bm k}^\prime}(E) = {\bm f}(E,{\bm k}) \delta_{{\bm k} {\bm k}^\prime},
\label{f}\\
&&
{\bm f}(E, {\bm k}) = \frac{ 2 \Delta  }{\Pi(E, {\bm k})} \, (\xi_{\bm k} {\bm \gamma}_{\bm k} - E \bm{h} - i {\bm \gamma}_{\bm k} \times \bm{h}),
\label{f_k}
\eea
where vector ${\bm f}(E,{\bm k})$ represents the triplet order parameter. 
Note that any weak SOC and Zeeman fields can induce a finite ${\bm f}$ - vector.
For this reason and in order to handle the upcoming calculations, 
we neglect ${\bm \gamma}_{\bm k}$ and $\bm{h}$ in the denominator, $\Pi(E, {\bm k})$, which takes then the form 

\bea
\Pi(E, {\bm k}) &\approx& (E^2 - \xi_{\bm k}^2 - \Delta^2)^2.
\label{Pi}
\eea
Above, $\xi_{\bm k}$ denotes the conduction-band dispersion given by 
$\xi_{\bm k} = (\varepsilon_1 + \lambda_1){\bm k}^2 - \mu$, where $\mu = -\varepsilon_0 - \lambda_0$ is the chemical potential measured 
from the bottom of the band.

As for the singlet amplitude $\hat{f}_0(E)$, it plays no role in the interface transport as long as the QAHI filters out pairs with zero spin.
For that, we choose the Fermi level at energy $E=0$ which lies in the gaps of both S and QAHI.
This situation is most suited for probing nonunitary states possessing a nonvanishing axial vector $i{\bm f}(0, {\bm k})\times {\bm f}^*(0, {\bm k})$.
In the case at hand,

\beq
i{\bm f}(0, {\bm k})\times {\bm f}^*(0, {\bm k})= 
\frac{8 \Delta^2 {\bm \gamma}_{\bm k}^2 \xi_{\bm k}}{\Pi(0, {\bm k})^2} \bm{h}.
\label{Axial}
\eeq 
We will argue that this axial vector acts as pair magnetization, causing a distinct magnetoresistance effect.

\subsection{Triplet magnetoresistance effect}
\label{LME}

The magnetoresistance effect is defined as the dependence of the interface resistance on an applied magnetic field. 
In our case, it is applied perpendicular to the interface, serving the twofold purpose of
stabilizing the magnetization in the QAHI and tuning Zeeman splitting $h$ in the S (\ref{SOC_Z}).
The magnetoresistance effect is caused by the dependence of transmission ${\cal T}_2$ (\ref{Tr_2}) on $h$
which comes primarily from the triplet ${\bm f}$ - vector.
To compute ${\cal T}_2(h)$, we need, apart from $\hat{A}$ (\ref{A_N_res}) and $\hat{F}$ (\ref{F_S}), the hopping matrix $\hat{t}$.
A widely used model is the hopping that depends neither on the spin nor on the orbital DOF. 
In this case, the hopping matrix can be written as the direct product of the coordinate-space matrix, $\hat{\zeta}$, and the unit matrices in spin and orbital spaces:
\beq
\hat{t} = \hat{\zeta} \sigma_0 s_0.
\label{t_diag} 
\eeq
Since all matrices in Eq. (\ref{Tr_2}) are now separable, we can trace out the spin and orbital DOF one after the other, using the usual algebra of Pauli matrices. 
As the calculation repeats that in \cite{GT17}, we just quote the result:

\bea
{\cal T}_2(E) &=& \frac{1}{4} 
{\rm Tr}\Bigl\{
\hat{\Gamma}(E) \bigl[ \hat{\bm{f}}(E) \times \bm{m} \bigr] \cdot \hat{\Gamma}^*(-E) \bigl[ \hat{\bm{f}}^\dagger(E) \times \bm{m} \bigr] 
\nonumber\\
&+& 
i\bm{m} \cdot \bigl[ \hat{\Gamma}(E) \hat{\bm{f}}(E) \times \hat{\Gamma}^*(-E) \hat{\bm{f}}^\dagger(E) \bigr] 
\Bigr\},
\label{Tr_2_coo}
\eea
where matrix $\hat{\Gamma}(E) = 2\pi \hat{\zeta}^\dagger \hat{\rho}(E) \hat{\zeta}$ yields (up to Planck's constant) single-particle hopping rates from or into the normal system, and the remaining trace operation pertains to a configuration space (i.e. position or ${\bm k}$ space).

Equation (\ref{Tr_2_coo}) is one of the main results and deserves a separate discussion.
It is valid for any triplet amplitude $\hat{\bm{f}}$ and unit magnetization $\bm{m}$.
The vanishing of transmission coefficient ${\cal T}_2$ for $\hat{\bm{f}}=0$ indicates transport of pure triplet pairs. 
In the S, the spin of the pair is perpendicular to $\hat{\bm{f}}$, while in the normal system it is parallel 
to magnetization direction $\bm{m}$. Therefore, the triplet transport is characterized by a noncollinear orientation of vectors $\hat{\bm{f}}$ and $\bm{m}$, 
which we indeed see in Eq. (\ref{Tr_2_coo}). Also, the noncollinearity of $\hat{\bm{f}}$ and $\bm{m}$ generally implies that 
the triplet transport is accompanied by torques on the magnetization. 

Furthermore, Eq. (\ref{Tr_2_coo}) accounts for nonunitary superconductivity with broken TRS. 
We see that the vector product of $\hat{\bm{f}}$ and $\hat{\bm{f}}^\dagger$ can indeed be ascribed to the pair magnetization, 
as it couples directly to the magnetization of the normal system. 
The magnetic coupling is a nonequilibrium transport effect which involves the pair hopping rates given by 
matrices $\hat{\Gamma}(E)$ and $\hat{\Gamma}^*(-E)$.

Besides, Eq. (\ref{Tr_2_coo}) provides a generalization of the two-particle transmission coefficient obtained in \cite{GT17}. 
The necessity of this generalization is dictated by several reasons. 
For instance, Ref. \cite{GT17} employs a single-orbital model (plus the spin), while the majority of quantum materials are multi-orbital systems.
In Eq. (\ref{Tr_2_coo}), the orbital DOF result in the prefactor of $1/4$. 
Such a substantial suppression of the transmission can be traced back to the fact that the edge mode is 
an equal-weight mixture of the conduction- and valence-band orbitals [see Eq. (\ref{A_N_res})]. 
Therefore, only half of the orbital state per electron can be transmitted from the QAHI into the S conduction band.
Alternatively, this can be explained by the mismatch of the orbital pseudospin, ${\bm s}=\frac{1}{2}[s_x, s_y, s_z]$,
which points in the $y$ and $z$ directions in the QAHI and S, respectively.
Another limitation of \cite{GT17} is the use of the translationally invariant Green's functions in ${\bm k}$ space.
In contrast, Eq. (\ref{Tr_2_coo}) does not invoke any translational invariance, allowing both position and ${\bm k}$ representations 
on an equal footing. 

To proceed, we take advantage of the 2D character of the QAHI and S layers,
allowing the evaluation of the trace in the basis of the in-plane ${\bm k}$ states as

\begin{widetext}
\bea
{\cal T}_2(E) = \frac{1}{4} \sum_{{\bm k}_1{\bm k}_2{\bm k}_3{\bm k}_4} \Gamma_{{\bm k}_1{\bm k}_2}(E) \Gamma^*_{{\bm k}_3{\bm k}_4}(-E)
\Bigl\{
\bigl[ \bm{f}_{{\bm k}_2{\bm k}_3}(E) \times \bm{m} \bigr] \cdot 
\bigl[ \bm{f}^*_{{\bm k}_1{\bm k}_4}(E) \times \bm{m} \bigr] 
+ 
i\bm{m} \cdot \bigl[ \bm{f}_{{\bm k}_2{\bm k}_3}(E)  \times \bm{f}^*_{{\bm k}_1{\bm k}_4}(E) \bigr] 
\Bigr\}.
\label{Tr_2_k}
\eea
\end{widetext}
Here,
${\bm f}_{{\bm k} {\bm k}^\prime}(E)$ is a diagonal matrix given by Eq. (\ref{f}), and 
$
\Gamma_{{\bm k}{\bm k}^\prime}(E) = 2\pi \sum_{{\bm k}_1{\bm k}_2} 
\zeta^*_{{\bm k}_1{\bm k}}\rho_{{\bm k}_1{\bm k}_2}(E)\zeta_{{\bm k}_2{\bm k}^\prime}
$
is the rate matrix, where
$\zeta_{{\bm k}{\bm k}^\prime}$ is the interlayer hopping matrix, and $\rho_{{\bm k}_1{\bm k}_2}(E)$ is the Fourier transform of Eq. (\ref{rho}) 
in the 2D ${\bm k}$ space.
Generally, the hopping matrix is not diagonal, neither is the spectral matrix $\rho_{{\bm k}_1{\bm k}_2}(E)$ because the QAHI 
has no translational invariance due to the boundary.
We can however adopt the widely used “broad-band” approximation, 
$\zeta_{{\bm k}{\bm k}^\prime} \approx \zeta = const$, making the calculation easier. 
In this approximation, the elements of the rate matrix are ${\bm k}$ - independent: 
$\Gamma_{{\bm k}{\bm k}^\prime}(E) \approx \Gamma(E) = 2\pi |\zeta|^2 \sum_{{\bm k}_1{\bm k}_2} \rho_{{\bm k}_1{\bm k}_2}(E) =
2\pi |\zeta|^2 {\cal N} \rho_{{\bm r} ={\bm r}^\prime=0}(E)$, where we used the Fourier transformation back to position space 
(${\cal N}$ is the number of unit cells in the normal layer). After some algebra involving Eqs. \ref{f_k} and \ref{Axial}, we find

\bea
{\cal T}_2(E,h) &=& \Delta^2 \Gamma(E) \Gamma(-E) \sum_{\bm k} 
\frac{ {\bm \gamma}^2_{\bm k} (\xi_{\bm k}  + {\bm m}\cdot \bm{h})^2}{\Pi(E,{\bm k})^2}
\label{Tr_2_h1}\\
&=& \Delta^2 \Gamma(E) \Gamma(-E) \sum_{\bm k} 
\frac{ {\bm \gamma}^2_{\bm k} (\xi_{\bm k}  +{\rm sgn}(m) h)^2}{\Pi(E,{\bm k})^2},
\label{Tr_2_h2}\quad\,\,\,
\eea
with ${\bm m} = {\rm sgn}(m) {\bm z}$ from Eq. (\ref{A_N_res}).
Finally, we define the zero-bias and -temperature electric conductance, $G(h) = \frac{e^2}{\pi\hbar}{\cal T}_2(0,h)$,  
resistance, $R(h)=1/G(h)$, and the relative change in the resistance

\bea
\frac{R(h) - R(0)}{R(h)} = \frac{G(0) - G(h)}{G(0)} = \frac{{\cal T}_2(0,0) - {\cal T}_2(0,h)}{{\cal T}_2(0,0)}.
\label{dR_def}
\eea
This ratio quantifies the triplet magnetoresistance effect discussed in detail below.

\section{Discussion and conclusions}

A striking feature of the triplet magnetoresistance effect is its dependence on the relative orientation of the magnetizations in the QAHI and S,
which is captured by the signs of $m$ and $h$ in Eq. (\ref{Tr_2_h2}). For parallel magnetizations, Zeeman energy $h$ has the same sign as $m$, i.e.

\beq
h = |h| {\rm sgn}(m),
\label{P}
\eeq 
and we recover magnetoresistance formula (\ref{dR}) announced in Introduction,  where
parameters $\alpha$ and $\beta$ depend on the Fermi energy and excitation gap in the S:  

\bea
\alpha\left( \frac{\mu}{\Delta} \right) &=& \int^\infty_{-\frac{\mu}{\Delta} } \frac{ \left(\frac{\mu}{\Delta} +x\right) dx}{(1+x^2)^4} /
                                                                \int^\infty_{-\frac{\mu}{\Delta} } \frac{ \left(\frac{\mu}{\Delta} +x\right)x^2dx}{(1+x^2)^4}
\label{alpha}\quad
\eea
and 
\bea
\beta\left( \frac{\mu}{\Delta} \right) &=& 2\int^\infty_{-\frac{\mu}{\Delta} } \frac{ \left(\frac{\mu}{\Delta} +x\right) xdx}{(1+x^2)^4} / 
                                                            \int^\infty_{-\frac{\mu}{\Delta} } \frac{ \left(\frac{\mu}{\Delta} +x\right)dx}{(1+x^2)^4}.
\label{beta}\quad
\eea
Asymptotically, $\alpha\approx 5$ and $\beta\approx \frac{2}{5} \frac{\Delta}{\mu}$ for $\mu\gg\Delta$.
The magnetoresistance behaviour plotted in Fig. \ref{dR_fig} is specific to the spin-triplet transport.
For weak Zeeman splitting, $|h| \ll \Delta$, the relative magnetoresistance varies linearly with $|h|$, 
which is a direct indicator of the nonunitary order parameter in the S. 
With increasing $\frac{\mu}{\Delta}$, the quadratic $h$ - dependence becomes visibly pronounced, 
although for small enough $h$ the magnetoresistance is always linear.

\begin{figure}[t]
\begin{center}
\includegraphics[width=90mm]{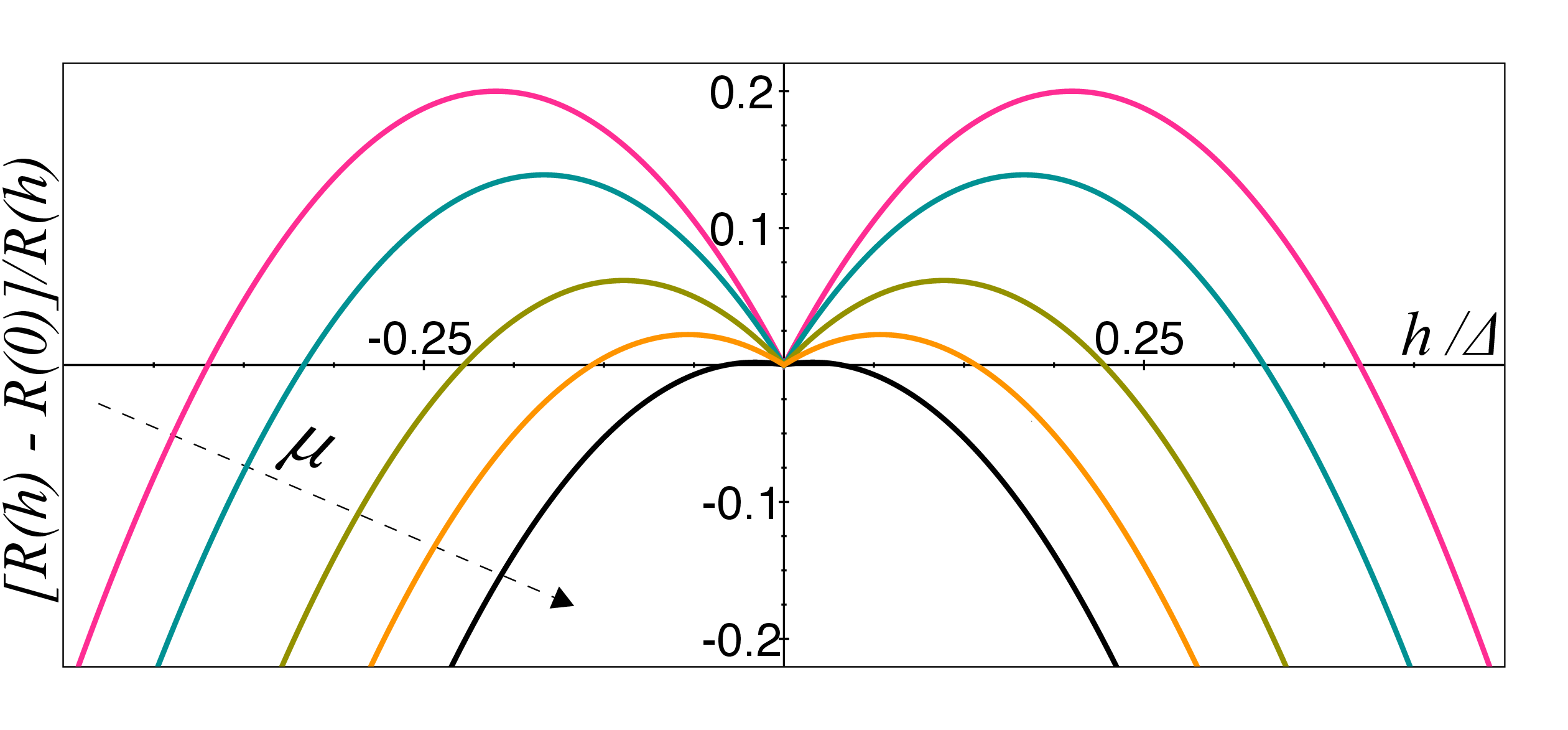}
\end{center}
\caption{
Interface magnetoresistance (\ref{dR_AP}) versus Zeeman splitting in the superconductor.
The arrow indicates an increasing chemical potential from $\mu = \Delta$ to $\mu = 10 \Delta$ (see also Eqs. \ref{alpha} and \ref{beta}).
}
\label{dR_AP_fig}
\end{figure}

The parallel orientation of the magnetizations is favoured because the magnetic field emerging from the QAHI naturally tends to magnetize the S in the same direction. This, however, does not exclude the case of the antiparallel magnetizations, e.g. if one of the materials has a negative $g$-factor.
In this case, the sign of Zeeman energy $h$ is opposite to ${\rm sgn}(m)$:
\beq
h = - |h| {\rm sgn}(m),
\label{AP}
\eeq 
and the resistance ratio in Eq. (\ref{dR_def}) takes the form 

\beq
\frac{R(h) - R(0)}{R(h)} = \alpha \frac{|h|}{\Delta} \left( \beta - \frac{|h|}{\Delta} \right),
\label{dR_AP}
\eeq
with same $\alpha$ and $\beta$. 
Unlike the parallel case, the low-field magnetoresistance is now positive, turning negative only at large enough $h$, as plotted in Fig. \ref{dR_AP_fig}.
Still, the linear magnetoresistance for $|h| \ll \Delta$ unmistakably signals the nonunitary superconductivity.

It is worth emphasizing that the above results pertain to an intraband nonunitary pairing in the spin space.  
Albeit not involved in the pairing, the band DOF affect the transport due to the mismatch of the “band pseudospins” 
in the QAHI and S, thereby reducing the pair transmission by the factor of 4. That is, 
the pair transport studied here is a multi-band problem in which both spin and band DOF need to be taken into account. 
Another specific of the model is the absence of the electron interaction in the triplet channel.
In this case, the nonunitary superconductivity is an emergent phenomenon due to the combined effect of Rashba and Zeeman interactions, 
allowing one to build a prototype theory of the linear triplet magnetoresistance effect.
The proposed formula for the pair transmission, Eq. (\ref{Tr_2_coo}), is however valid beyond that particular model. 
Also for intrinsic nonunitary states, the Onsager symmetry requires that an external magnetization couples to the odd powers of a magnetic field in the superconductor 
via an axial vector $i\hat{\bm f} \times \hat{\bm f}^\c$. 
The presented theory can therefore be implemented for transport identification of TRS - breaking nonunitary states in 
intrinsic multi-band superconductors as well.

Lastly, the use of the “broad-band” approximation in Eq. (\ref{Tr_2_k}) merits brief comment.
The underlying physics is clearly independent of this approximation 
because for any interface the sum in Eq. (\ref{Tr_2_k}) contains the terms proportional to 
axial vector $i\hat{\bm f} \times \hat{\bm f}^\c$, the hallmark of a nonunitary state.
The proposed calculation of the two-particle transmission can naturally be extended to other interesting interface models,
which will be a subject of future studies.

\acknowledgments
The author thanks Thilo Kopp, Arno Kampf and Juan Carlos Cuevas for their valuable comments.
This work was partially supported by the German Research Foundation (DFG) through TRR 80.

\appendix
\section{One- and two-particle currents from NEGFs}
\label{A}

Since the interface hopping in Hamiltonian (\ref{H_C}) is treated as a potential, we obtain a Lippmann-Schwinger-like equation for NEGF $\breve{G}(E)$ and
can therefore write the solution in the form of an iterative series:

\bea
\breve{G}(E) &=& \breve{g}(E) + \breve{g}(E) \breve{V} \breve{g}(E) + \breve{g}(E) \breve{V} \breve{g}(E)\breve{V} \breve{g}(E) 
\nonumber\\ 
&+& 
\breve{g}(E) \breve{V} \breve{g}(E)\breve{V} \breve{g}(E)\breve{V} \breve{g}(E) +...
\label{Iter_sol}
\eea
Here, $\breve{g}(E)$ is the Green's function in the absence of the interface hopping, and $\breve{V}$ is the hopping matrix in the NEGF representation:

\beq
\breve{V} = \left[
\begin{matrix}
+\hat{V} & 0 \\
0 & - \hat{V}
\end{matrix}
\right],
\label{V_Keldysh}
\eeq
where signs $\pm$ reflect the opposite directions of the time integration on the upper and lower branches of the Keldysh contour.
The matrix structure of $\breve{g}(E)$ is analogous to that in Eq. (\ref{G_NE}),
so, by doing the matrix multiplication in Eq. (\ref{Iter_sol}), one can obtain the required (greater) function $\hat{G}^{>}(E)$ to any order in hopping:

\beq
\hat{G}^{>}(E) = ...+\hat{G}^>_1(E) + ... + \hat{G}^>_2(E) +...,
\eeq
where labels $1, 2, ...$ refer to one-particle, two-particle and higher-order many-particle hoppings. 
In the following, we focus on the one- and two-particle currents. 

\subsection{One-particle current}
\label{One}

It arises from the second term in the series in Eq. (\ref{Iter_sol}).
The expression for $\hat{G}^>_1(E)$ is

\bea
\hat{G}^>_1(E) = \frac{1}{2}
&
\left(
\left[ \hat{g}^K(E) - \hat{g}^A(E) \right] \hat{V} \hat{g}^A(E)+
\right.
&
\nonumber\\
&
\left.
\hat{g}^R(E) \hat{V} \left[ \hat{g}^K(E) + \hat{g}^R(E) \right]
\right),
&
\label{G_1}
\eea
which yields the single-particle current [see Eq. (\ref{I_G})]

\bea
{\cal I}_1 = - \frac{e}{8\pi\hbar} 
\int
&
{\rm Tr}\Bigl[
\nu_3\tau_3i_0 \hat{V} \hat{g}^K(E) \hat{V} \hat{g}^A(E)+
&
\nonumber\\
&
\nu_3\tau_3i_0\hat{V} \hat{g}^R(E) \hat{V} \hat{g}^K(E)
\Bigr]dE.
&
\label{I_1}
\eea
The integrals with $\hat{g}^A(E)\hat{V} \hat{g}^A(E)$ and $\hat{g}^R(E)\hat{V} \hat{g}^R(E)$ vanish
due to the analytical properties of the retarded and advanced functions. 

Next, we trace over SN space, using Eq. (\ref{V_ab}) for $\hat{V}$ and 
the diagonal Green's functions 

\beq
\hat{g}^{K,A,R}
= \left[
\begin{matrix}
\hat{g}^{K,A,R}_{_S} & 0 \\
0 & \hat{g}^{K,A,R}_{_N}
\end{matrix}
\right],
\label{g}
\eeq
with the following result

\bea
{\cal I}_1 = \frac{e}{8\pi\hbar} && \int 
{\rm Tr}\Bigl(
\tau_3i_0 \hat{T} \hat{g}^K_{_S}(E) \hat{T}^\c [\hat{g}^A_{_N}(E) - \hat{g}^R_{_N}(E)]-
\nonumber\\
&&
\tau_3i_0 \hat{T} [\hat{g}^A_{_S}(E) - \hat{g}^R_{_S}(E)] \hat{T}^\c \hat{g}^K_{_N}(E)
\Bigr)
dE.
\label{I_1_SN}
\eea
Further, we insert $\hat{T}$ from Eq. (\ref{T_ph}) and $\hat{g}^{K,A,R}_{_N}(E)$ given (in ph space) by

\beq
\hat{g}^{A,R}_{_N}(E)
= \left[
\begin{matrix}
\hat{G}^{A,R}_{_N}(E) & 0 \\
0 & -\hat{G}^{A,R}_{_N}(-E)^*
\end{matrix}
\right]
\label{g_AR}
\eeq
and

\beq
\hat{g}^K_{_N}(E)
= \left[
\begin{matrix}
\hat{G}^K_{_N}(E) & 0 \\
0 & \hat{G}^K_{_N}(-E)^*
\end{matrix}
\right],
\label{g_K}
\eeq
where

\bea
\hat{G}^K_{_N}(E) = -2\pi i \hat{A}_{_N}(E)[1-2f_{_N}(E)],
\label{G^K_N}
\eea
with the spectral function

\bea
\hat{A}_{_N}(E) = \frac{\hat{G}^A_{_N}(E) - \hat{G}^R_{_N}(E)}{2\pi i}
\label{A_N}
\eea
and the fermion distribution function $f_{_N}(E)$.
Since $\hat{T}$ and $\hat{g}^{K,A,R}_{_N}(E)$ are diagonal,
only the diagonal parts of matrices $\hat{g}^{K,A,R}_{_S}(E)$ contribute to the trace over ph space. 
These have the same structure as in Eqs. (\ref{g_AR}) -- (\ref{A_N}), where label N is replaced with S. 
The one-particle current can then be cast in the form

\bea
{\cal I}_1 &=& \frac{e}{4\pi\hbar}
\int {\cal T}_1(E)[f_{_N}(E) - f_{_S}(E)]dE
\label{I_1_p}\\
&+& 
\frac{e}{4\pi\hbar}
\int 
{\cal T}^*_1(-E)[f_{_N}(-E) - f_{_S}(-E)]
dE\quad
\label{I_1_h}\\
&=& 
\frac{e}{2\pi\hbar} \int {\cal T}_1(E) [f_{_N}(E) - f_{_S}(E)],
\nonumber
\eea
with the one-particle transmission coefficient

\beq
{\cal T}_1(E) = (2\pi)^2{\rm Tr}\Bigl[
\hat{t}^\c \hat{A}_{_N}(E)\hat{t} \hat{A}_{_S}(E) 
\Bigr].
\label{Tr_1}
\eeq
Since the latter is real-valued, the particle and hole contributions to ${\cal I}_1$ are identical [cf. Eqs. (\ref{I_1_p}) and (\ref{I_1_h})].

\subsection{Two-particle current}
\label{Two}

The two-particle current is related to the correction $\hat{G}^>_2(E)$ from the fourth term in the series in Eq. (\ref{Iter_sol}).
The direct matrix multiplication yields 

\bea
\hat{G}^>_2(E) = \frac{1}{2}
&
\Bigl[ 
\hat{g}^K(E) \hat{V} \hat{g}^A(E) \hat{V} \hat{g}^A(E) \hat{V} \hat{g}^A(E)
&
\nonumber\\
&
+ \hat{g}^R(E) \hat{V} \hat{g}^K(E) \hat{V} \hat{g}^A(E) \hat{V} \hat{g}^A(E)
&
\nonumber\\
&
+ \hat{g}^R(E) \hat{V} \hat{g}^R(E) \hat{V} \hat{g}^K(E) \hat{V} \hat{g}^A(E)
&
\nonumber\\
&
+ \hat{g}^R(E) \hat{V} \hat{g}^R(E) \hat{V} \hat{g}^R(E) \hat{V} \hat{g}^K(E)
&
\nonumber\\
&
+ \hat{g}^R(E) \hat{V} \hat{g}^R(E) \hat{V} \hat{g}^R(E) \hat{V} \hat{g}^R(E)
&
\nonumber\\
&
- \hat{g}^A(E) \hat{V} \hat{g}^A(E) \hat{V} \hat{g}^A(E) \hat{V} \hat{g}^A(E)
\Bigr].
&
\label{G_2}
\eea
The last two terms do not contribute to current (\ref{I_G_E}) for the same reason as in the one-particle case.
Hence, the two-particle current

\bea
{\cal I}_2 &=& - \frac{e}{8\pi\hbar} 
\int
{\rm Tr}\Bigl\{
\nu_3\tau_3i_0 \Bigl[
\nonumber\\
& &
\hat{V} \hat{g}^K(E) \hat{V} \hat{g}^A(E) \hat{V} \hat{g}^A(E) \hat{V} \hat{g}^A(E) +
\nonumber\\
& &
\hat{V} \hat{g}^R(E) \hat{V} \hat{g}^K(E) \hat{V} \hat{g}^A(E) \hat{V} \hat{g}^A(E) +
\nonumber\\
& &
\hat{V} \hat{g}^R(E) \hat{V} \hat{g}^R(E) \hat{V} \hat{g}^K(E) \hat{V} \hat{g}^A(E) +
\nonumber\\
& & 
\hat{V} \hat{g}^R(E) \hat{V} \hat{g}^R(E) \hat{V} \hat{g}^R(E) \hat{V} \hat{g}^K(E)
\Bigr] \Bigr\}dE.
\quad
\label{I_2}
\eea
To calculate the trace in SN space, we again use Eq. (\ref{V_ab}) for $\hat{V}$ and Eq. (\ref{g}) for $\hat{g}^{K,A,R}$ and find

\bea
{\cal I}_2 &=& - \frac{e}{8\pi\hbar} 
\int
{\rm Tr}\Bigl\{
\tau_3i_0 \Bigl[
\nonumber\\
& &
\hat{T} \hat{g}^K_{_S}(E) \hat{T}^\c \hat{g}^A_{_N}(E) \hat{T} \hat{g}^A_{_S}(E) \hat{T}^\c [\hat{g}^A_{_N}(E) - \hat{g}^R_{_N}(E)]+
\nonumber\\
& &
\hat{T} \hat{g}^R_{_S}(E) \hat{T}^\c \hat{g}^K_{_N}(E) \hat{T} \hat{g}^A_{_S}(E) \hat{T}^\c [\hat{g}^A_{_N}(E) - \hat{g}^R_{_N}(E)]+
\nonumber\\
& &
\hat{T} \hat{g}^R_{_S}(E) \hat{T}^\c \hat{g}^R_{_N}(E) \hat{T} \hat{g}^K_{_S}(E) \hat{T}^\c [\hat{g}^A_{_N}(E) - \hat{g}^R_{_N}(E)]+
\nonumber\\
& &
\hat{T} \hat{g}^R_{_S}(E) \hat{T}^\c \hat{g}^R_{_N}(E) \hat{T} \hat{g}^R_{_S}(E) \hat{T}^\c \hat{g}^K_{_N}(E) -
\nonumber\\
& &
\hat{T} \hat{g}^A_{_S}(E) \hat{T}^\c \hat{g}^A_{_N}(E) \hat{T} \hat{g}^A_{_S}(E) \hat{T}^\c \hat{g}^K_{_N}(E)
\Bigr] \Bigr\}dE.
\quad
\label{I_2_SN}
\eea

Generally, the two-particle current is a subleading correction. However, under conditions 

\beq
\hat{g}^A_{_S}(E) = \hat{g}^R_{_S}(E) \equiv \hat{g}_{_S}(E), \qquad \hat{g}^K_{_S}(E) =0,
\label{Conditions_I_2}
\eeq 
the one-particle current in Eq. (\ref{I_1_SN}) is suppressed, and ${\cal I}_2$ becomes the leading order.
This is the case if S has a single-particle gap $\Delta$ and the energies of interest are much smaller than $\Delta$. 
In the following, we specifically focus on this regime. It turns out that under conditions (\ref{Conditions_I_2}) 
only the off-diagonal part of $\hat{g}_{_S}(E)$ contributes to the trace in ph space. 
We therefore use

\beq
\hat{g}_{_S}(E)
= \left[
\begin{matrix}
0 & \hat{F}_{_S}(E) \\
\hat{F}^\c_{_S}(E) & 0
\end{matrix}
\right]
\label{g_S}
\eeq
along with Eq. (\ref{T_ph}) for $\hat{T}$ and Eqs. (\ref{g_AR}) and (\ref{g_K}) for $\hat{g}^{K,A,R}_{_N}(E)$
to find

\bea
&&
{\cal I}_2 = \frac{e}{4\pi\hbar} 
\int
{\rm Tr}\Bigl[
\label{I_2_spin}\\
&&
\hat{t}^{\c *} \hat{G}^K_{_N}(-E)^* \hat{t}^* \hat{F}^\c_{_S}(E) \hat{t}^\c [\hat{G}^A_{_N}(E) - \hat{G}^R_{_N}(E)] \hat{t} \hat{F}_{_S}(E) +
\nonumber\\
&&
\hat{t}^\c \hat{G}^K_{_N}(E) \hat{t} \hat{F}_{_S}(E) \hat{t}^{\c *} [\hat{G}^A_{_N}(-E) - \hat{G}^R_{_N}(-E)]^* \hat{t}^* \hat{F}^\c_{_S}(E)
\Bigr] dE,
\nonumber
\eea
where $\hat{F}_{_S}(E)$ is the anomalous (condensate) Green's function. 
Finally, in view of Eqs. (\ref{G^K_N}) and (\ref{A_N}), the two-particle current can be cast into the form presented in the main text 
[see Eqs. (\ref{I_2_final}) and (\ref{Tr_2})]. To ease the notation, we omit there subscripts $S$ and $N$ at the Green’s functions.


\begin{thebibliography}{99}

\bibitem{Keimer17}
B. Keimer and J. Moore, The physics of quantum materials, Nature Phys. {\bf 13}, 1045 (2017). 

\bibitem{Giustino20}
F. Giustino et al, The 2021 quantum materials roadmap, J. Phys. Mater. {\bf 3}, 042006 (2020).

\bibitem{Liu16}
C.-X. Liu, S.-C. Zhang, and X.-L. Qi, 
The quantum anomalous Hall effect: theory and experiment,
Annu. Rev. Condens. Matter Phys. {\bf 7}, 301 (2016).

\bibitem{Armitage18}
N. Armitage, E. Mele, and A. Vishwanath, Weyl and Dirac semimetals in three-dimensional solids, Rev. Mod. Phys. {\bf 90}, 015001 (2018).

\bibitem{Choi17}
W. Choi, N. Choudhary, G. H. Han, J. Park, D. Akinwande, and Y. H. Lee,
Recent development of two-dimensional transition metal dichalcogenides and their applications,
Materials Today {\bf 20}, 116 (2017).

\bibitem{Tanaka12}
Y. Tanaka, M. Sato, and N. Nagaosa, 
Symmetry and topology in superconductors -- odd-frequency pairing and edge states,
J. Phys. Soc. Jpn. {\bf 81}, 011013 (2012).

\bibitem{Sato17}
M. Sato and Y. Ando, Topological superconductors: a review, Rep. Prog. Phys. {\bf 80}, 076501 (2017).

\bibitem{Culcer20}
D. Culcer, A.C. Keser, Y. Li and G. Tkachov, Transport in two-dimensional topological materials: recent
developments in experiment and theory, 2D Mater. {\bf 7}, 022007 (2020).

\bibitem{Ghosh20}
S. K. Ghosh, M. Smidman, T. Shang, J. F. Annett, A. D. Hillier, J. Quintanilla J, and H. Yuan, 
Recent progress on superconductors with time-reversal symmetry breaking,
J. Phys.: Condens. Matter. {\bf 33}, 033001 (2020).

\bibitem{Nakai21}
R. Nakai, K. Nomura, and Y. Tanaka,
Edge-induced pairing states in a Josephson junction through a spin-polarized quantum anomalous Hall insulator,
Phys. Rev. B {\bf 103}, 184509 (2021).

\bibitem{Cheng21}
Q. Cheng, Q. Yan, and Q.-F. Sun,
Spin-triplet superconductor -- quantum anomalous Hall insulator -- spin-triplet superconductor Josephson junctions: 
$0-\pi$, $\phi_0$ phase, and switching effects,
Phys. Rev. B {\bf 104}, 134514 (2021).

\bibitem{Ohashi21}
R. Ohashi, S. Kobayashi, and Y. Tanaka,
Possible topological phases in quantum anomalous Hall insulator/unconventional superconductor hybrid systems,
Phys. Rev. B {\bf 104}, 134518 (2021).

\bibitem{Ramires22}
A. Ramires, Nonunitary superconductivity in complex quantum materials,
J. Phys.: Condens. Matter {\bf 34}, 304001 (2022).


\bibitem{Blonder82}
G. E. Blonder, M. Tinkham, and T. M. Klapwijk, 
{\em Transition from metallic to tunneling regimes in superconducting microconstrictions: Excess current, charge imbalance, and supercurrent conversion}, 
Phys. Rev. B {\bf 25}, 4515 (1982).

\bibitem{Cuevas96}
J. C. Cuevas, A. Mart{\'i}n-Rodero, and A. Levy Yeyati, 
Hamiltonian approach to the transport properties of superconducting quantum point contacts, 
Phys. Rev. B {\bf 54}, 7366 (1996).

\bibitem{GT17}
G. Tkachov, Magnetoelectric Andreev effect due to proximity-induced nonunitary triplet superconductivity in helical metals,
Phys. Rev. Lett. {\bf 118}, 016802 (2017).

\bibitem{Hillier09}
A. D. Hillier, J. Quintanilla, and R. Cywinski, 
Evidence for time-reversal symmetry breaking in the noncentrosymmetric superconductor LaNiC$_2$,
Phys. Rev. Lett. {\bf 102}, 117007 (2009).

\bibitem{Hillier12}
A. D. Hillier, J. Quintanilla, B. Mazidian, J. F. Annett, and R. Cywinski,
Nonunitary triplet pairing in the centrosymmetric superconductor LaNiGa$_2$,
Phys. Rev. Lett. {\bf 109}, 097001 (2012).

\bibitem{Lado19}
J. L. Lado and M. Sigrist, 
Detecting nonunitary multiorbital superconductivity with Dirac points at finite energies, 
Phys. Rev. Research {\bf 1}, 033107 (2019).

\bibitem{Shang22}
T. Shang, S. K. Ghosh, M. Smidman et al., Spin-triplet superconductivity in Weyl nodal-line semimetals, 
npj Quantum Mater. {\bf 7}, 35 (2022).

\bibitem{Ghosh22}
S. K. Ghosh, P. K. Biswas, C. Xu, B. Li, J. Z. Zhao, A. D. Hillier, and X. Xu,
Time-reversal symmetry breaking superconductivity in three-dimensional Dirac semimetallic silicides,
Phys. Rev. Research {\bf 4}, L012031 (2022).

\bibitem{Wolf22}
T. M. R. Wolf, M. F. Holst, M. Sigrist, and J. L. Lado,
Nonunitary multiorbital superconductivity from competing interactions in Dirac materials,
Phys. Rev. Research {\bf 4}, L012036 (2022).

\bibitem{Sigrist91}
M. Sigrist and K. Ueda, 
Phenomenological theory of unconventional superconductivity,
Rev. Mod. Phys. {\bf 63}, 239 (1991).

\bibitem{Honerkamp98}
C. Honerkamp and M. Sigrist, 
Andreev reflection in unitary and non-unitary triplet states,
J. Low Temp. Phys. {\bf 111}, 895 (1998).

\bibitem{Linder07}
J. Linder, M. S. Gronsleth, and A. Sudbo, 
Conductance spectra of ferromagnetic superconductors: 
quantum transport in a ferromagnetic metal/non-unitary ferromagnetic superconductor junction, 
Phys. Rev. B {\bf 75}, 054518 (2007).

\bibitem{GT19}
G. Tkachov, 
Probing the magnetoelectric effect in noncentrosymmetric superconductors by equal-spin Andreev tunneling, 
J. Phys.: Condens. Matter {\bf 31}, 055301 (2019).

\bibitem{Kashiwaya00}
S. Kashiwaya and Y. Tanaka, Tunnelling effects on surface bound states in unconventional superconductors, 
Rep. Prog. Phys. {\bf 63}, 1641 (2000). 

\bibitem{Shi19}
L.-K. Shi and J. C. W. Song, 
Symmetry, spin-texture, and tunable quantum geometry in a WTe$_2$ monolayer,
Phys. Rev. B {\bf 99}, 035403 (2019).

\bibitem{GT22}
G. Tkachov, {\it Topological Quantum Materials: Concepts, Models, and Phenomena} (Jenny Stanford Publishing, New York, 2022).

\bibitem{Loder13}
F. Loder, A. P. Kampf, and T. Kopp, Superconductivity with Rashba spin-orbit coupling and magnetic field, 
J. Phys.: Condens. Matter {\bf 25}, 362201 (2013).

\bibitem{Loder15}
F. Loder, A. Kampf, and T. Kopp, Route to topological superconductivity via magnetic field rotation. Sci. Rep. {\bf 5}, 15302 (2015).


\end{thebibliography}
\end{document}